\documentclass[]{spie}  

\usepackage{amsmath,amsfonts,amssymb}
\usepackage{graphicx}
\usepackage{tikz}
\usepackage{subcaption}
\usepackage[percent]{overpic}
\usepackage[colorlinks=true, allcolors=blue]{hyperref}
\usepackage{subcaption}
\usepackage{float}
\usepackage{bm}

\title{Modal estimation of high-order wavefront drifts on coronagraphic testbeds for high-contrast direct imaging of exoplanets}

\author[a]{Saikrishna Manojkumar}
\author[b,c]{Susan F. Redmond}
\author[d]{Leonid Pogorelyuk}
\author[a]{Christine L. Page}
\author[a]{Ajay S. Gill}
\author[e]{Laurent Pueyo}
\author[f]{Emiel H. Por}
\author[g]{Iva Laginja}
\author[e]{Rapha\"el Pourcelot}
\author[e]{Bryony F. Nickson}
\author[h]{Ananya Sahoo}
\author[e]{Meiji M. Nguyen}
\author[e]{R\'emi Soummer}
\author[e]{Marshall D. Perrin}
\author[i]{Bijan Nemati}
\author[j]{Jeremy N. Kasdin}
\author[a]{Kerri Cahoy}

\affil[a]{Massachusetts Institute of Technology, 77 Massachusetts Ave, Cambridge, MA 02139, USA}
\affil[b]{California Institute of Technology, 1216 E California Blvd, Pasadena, CA 91125, USA}
\affil[c]{Jet Propulsion Laboratory, 4800 Oak Grove Dr, Pasadena, CA 91011, USA}
\affil[d]{Rensselaer Polytechnic Institute, 110 8th St, Troy, NY 12180, USA}
\affil[e]{Space Telescope Science Institute, 3700 San Martin Drive, Baltimore, MD 21218, USA}
\affil[f]{University of California, Santa Cruz, 1156 High St Santa Cruz, CA 95064, USA}
\affil[g]{LESIA, Observatoire de Paris, Université PSL, Sorbonne Université, Université Paris Cité, CNRS, 5 pl. Jules Janssen, 92195 Meudon, France}
\affil[h]{Lowell Center for Space Science and Technology, University of Massachusetts, 600 Suffolk Street Suite 315, Lowell, MA 01854, USA}
\affil[i]{The University of Alabama in Huntsville, 301 Sparkman Dr NW, Huntsville, AL 35899, USA}
\affil[j]{Princeton University, Olden Street, Princeton, NJ 08544, USA}

\authorinfo{Further author information: (Send correspondence to S.M.)\\S.M.: E-mail: saimanoj@mit.edu}

\begin{document} 
\maketitle
\begin{abstract}
 Long-term wavefront stability will be critical for the upcoming Habitable Worlds Observatory which aims to directly image exo-Earths. Mechanical and thermal instabilities induce slow, high-order wavefront errors which can degrade the contrast and in turn, the scientific yield of the observatory. Dark zone maintenance (DZM) has been proposed as a high-order wavefront sensing and control strategy to maintain contrast despite wavefront error drifts. This approach uses an Extended Kalman Filter (EKF) to estimate the electric field at each pixel in the dark zone of the science camera. However, relying on individual pixel information to estimate wavefront errors can discard information stored in the underlying wavefront. This is a fundamental limitation on the best achieved contrast compared to the model-based theoretical bounds derived from photon noise. An alternative EKF can incorporate more wholistic information by estimating the drifting wavefront modes present in the system. In this work we analyze the performance of this Modal EKF on the High-contrast Imager for Complex Aperture Telescopes (HiCAT) simulator at the Space Telescope Science Institute. A parameter scan is performed across key variables: deformable mirror (DM) drift, the direct photon rate at the telescope aperture governing the mean number of photons in the dark zone (DZ), and the dither magnitude used to provide phase diversity. While prior studies explored a range of dither magnitudes, this work introduces a covariance-based approach within the Extended Kalman Filter (EKF) framework to optimally determine the probe magnitude required. Results from the modal EKF demonstrate a 15-fold improvement in contrast relative to the single-pixel DZM experiments, reducing the gap to within a factor of 2–3 of the theoretical bounds.
\end{abstract}


\section{INTRODUCTION}
\label{sec:intro}  
\subsection{Current Status of Direct Imaging with Space Telescopes}\label{sec:direct_imaging}
The Habitable Worlds Observatory (HWO), scheduled for launch in the 2040s, is envisioned as the next flagship large-aperture observatory tasked with directly imaging Earth-like exoplanets, as recommended by the 2020 Decadal Survey on Astronomy and Astrophysics. Imaging exoplanets in the habitable zone, where liquid water may exist, requires suppressing starlight that is approximately~$10^{10}$ times brighter than the planetary signal \cite{stark_exoearth_2019}. Coronagraphs, optical systems composed of a focal-plane mask to occult starlight and a pupil-plane mask to mitigate diffraction, have been demonstrated by NASA to achieve contrasts of $4\times10^{-10}$ when combined with two deformable mirrors (DMs) \cite{seo_testbed_2019}. This enables the formation of a high-contrast region in the image, known as the dark zone (DZ), by compensating for static aberrations in the telescope system.

The Nancy Grace Roman Space Telescope (RST) will provide the first in-space demonstration of this technology, targeting contrasts of~$10^{-8}$ with its coronagraph. In Roman’s observing strategy, a bright reference star is used to generate the DM commands needed to suppress aberrations and establish a DZ before slewing to the science target. However, thermal and mechanical instabilities degrade the DZ over time, requiring repeated slews back to the reference star to refresh DM commands \cite{kasdin_nancy_2020}. This approach reduces the effective observing time and is not practical for HWO, which demands a higher and more stable contrast over long durations.

To meet these requirements, high-order wavefront sensing and control (HOWFSC) techniques are needed. HOWFSC enables continuous observation of the science target while simultaneously correcting quasi-static aberrations, which manifest as speckles in the image~\cite{racine_speckle_1999}. One such technique is Dark Zone Maintenance (DZM), which uses science exposures directly as measurements to estimate and null aberrations to maintain high contrast in the DZ \cite{pogorelyuk_dark_2019, redmond_implementation_2022}.

\subsection{Approaches to Maintain Contrast in the Dark Zone}\label{subsec:appraoches}
Several algorithms have been developed to maintain contrast in the dark zone, including Linear Dark Field Control (LDFC) \cite{miller_spatial_2017} and implicit electric field conjugation (iEFC)\cite{haffert_implicit_2023} among others. DZM is one such approach, and has consistently demonstrated the ability to sustain contrast in the DZ despite continuous disturbances acting on the system. 

In DZM, science exposures are used directly as measurements for the Extended Kalman Filter (EKF), which serves as the primary estimator of the open-loop electric field at every pixel of the focal plane. The EKF output is then passed to the Electric Field Conjugation (EFC) controller, which combines the electric field estimate with the system Jacobian to compute DM commands that null the field via least-squares minimization \cite{giveon_broadband_2007}. EFC is also widely used to dig the DZ when paired with pair-wise probing (PWP), which estimates the electric field by differencing images taken with positive and negative probe commands applied to the DMs \cite{giveon_pair-wise_2011}. While the PWP and EFC combination has been shown to effectively create high-contrast dark zones, the large number of exposures required makes it inefficient for contrast maintenance during science observations.

\subsubsection{Contrast Limiting Parameters}\label{sec:contrast_limit_factors}
One of the primary metrics for evaluating contrast maintenance algorithms is their performance under contrast-limiting factors. These factors can be broadly classified into environmental and wavefront sensing and control (WFSC) variables. Environmental factors arise from the telescope and its operating environment, while WFSC factors are tied to the specific algorithm being applied.

Space telescope contrast is limited by a range of environmental effects that manifest across spatial orders~\cite{coyle_large_2019}. A major quasi-static disturbance is deformable mirror (DM) actuator drift, driven by thermal and mechanical instabilities. These drifts occur in the high-spatial-order regime, making them difficult to detect~\cite{mejia_prada_high-contrast_2019}. Prior work has characterized such drifts as random-walk processes with a power spectral density decaying as $f^{-2}$~\cite{pogorelyuk_effects_2020}. This allows the DM drift progression to be modeled as
\begin{equation}
\mathbf{u}_{k+1}^{drift} = \mathbf{u}_k^{drift} + \mathcal{N}(0, \sigma_{drift}^2 \mathbf{I}) ,
\end{equation}
where $\mathbf{u}_k^{drift}$ is the DM drift at iteration $k$, $\sigma_{drift}^2$ is the variance of the normal distribution from which drift commands are drawn, and $\mathbf{I}$ is the identity matrix. 

Another key environmental limitation is photon rate in coronagraphic imaging, particularly for dim target stars where system throughput is already constrained. The limited flux introduces Poisson (shot) noise at the detector~\cite{nemati_photon_2020}. Under low-light conditions this noise can dominate, reducing the effectiveness of algorithms that rely on coronagraph images for electric field estimation. Following Redmond et al. 2022~\cite{redmond_dark_2022}, this effect can be simulated by prescribing a direct photon rate at the telescope aperture.

The low-photon imaging process begins with a direct image (Lyot stop removed) taken at long exposure to ensure shot noise dominates over detector effects such as dark current and read noise. Summing the image yields a total photon rate $F$, which is compared to the prescribed rate $F_{target}$ to define a scaling factor
\begin{equation}
\gamma = \frac{F_{target}}{F} ,
\end{equation}
used to scale coronagraphic images, $I_{coron}$. Poisson noise is then injected as
\begin{equation}
I_{noisy} = \frac{\text{round}(P((\gamma I_{hicat} + \mu_{I})t_{exposure}))}{t_{exposure}} ,
\end{equation}
where $P$ denotes the Poisson distribution, $\mu_I$ is the dark current, and $t_{exposure}$ is the exposure time. From this prescription, the key metric reported is the mean number of photons per pixel in the DZ, $\kappa$~\cite{manojkumar_parameter_2024}.

On the WFSC side, the primary factor considered is the dither magnitude, which provides the phase diversity required by focal-plane sensing techniques to estimate the electric field~\cite{riggs_integrated_2016}. The dither magnitude, $\sigma_{dither}$, generates commands according to
\begin{equation}
\mathbf{u}_{k}^{dither} = \mathcal{N}(0, \sigma_{dither}^2 \mathbf{I}) .
\end{equation}
In DZM, the EKF, referred to as the single-pixel EKF, typically prescribes a dither magnitude $5\times$–$10\times$ larger than the drift magnitude to ensure sufficient phase diversity at each iteration~\cite{manojkumar_parameter_2024}. By contrast, the modal EKF described later requires significantly smaller dither amplitudes, and the use of stochastic gradient descent enables an optimal dither magnitude to be identified adaptively.

\subsubsection{DZM Previous Work}

DZM has been demonstrated on the in-air High-contrast Imager for Complex Aperture Telescopes (HiCAT) testbed at the Space Telescope Science Institute (STScI)~\cite{ndiaye_high-contrast_2019, soummer_high-contrast_2019, soummer_high-contrast_2022, soummer_high-contrast_2024}. These demonstrations, reported in Redmond et al.~2021~\cite{redmond_dark_2021} and Redmond et al. 2022~\cite{redmond_dark_2022}, raised the Technology Readiness Level (TRL) of the algorithm to 3. To evaluate its potential for future implementation, however, it is necessary to probe the performance limits of DZM under contrast-limiting factors and compare results to theoretical bounds derived from the photon noise and the information limit of the DZ~\cite{pogorelyuk_information-theoretical_2021}. This was pursued through a parameter scan, described in Manojkumar et al. 2024~\cite{manojkumar_parameter_2024}, which benchmarked DZM by comparing theoretical bounds, hardware results, and simulation results obtained using the \texttt{catkit2} simulator~\cite{noauthor_spacetelescopecatkit2_2024}.

The scan examined three primary parameters: (1) drift magnitude applied to the DMs, spanning $\sigma_{drift} = 0.03$–$3~\text{nm}/\sqrt{\text{iteration}}$; (2) direct photon rate at the primary mirror, ranging from $1.85 \times 10^6$ to $1.85 \times 10^8$ photons per second, corresponding to initial DZ photon levels of $\kappa = 0.15$–$1.98$ photons per pixel; and (3) dither magnitude, tested as multiples of the drift magnitude ($5\times$, $10\times$, $15\times$, and $20\times$).

\begin{figure}[htb]
    \centering
    \begin{subfigure}[b]{0.4\textwidth}
        \centering
        \includegraphics[width=\textwidth]{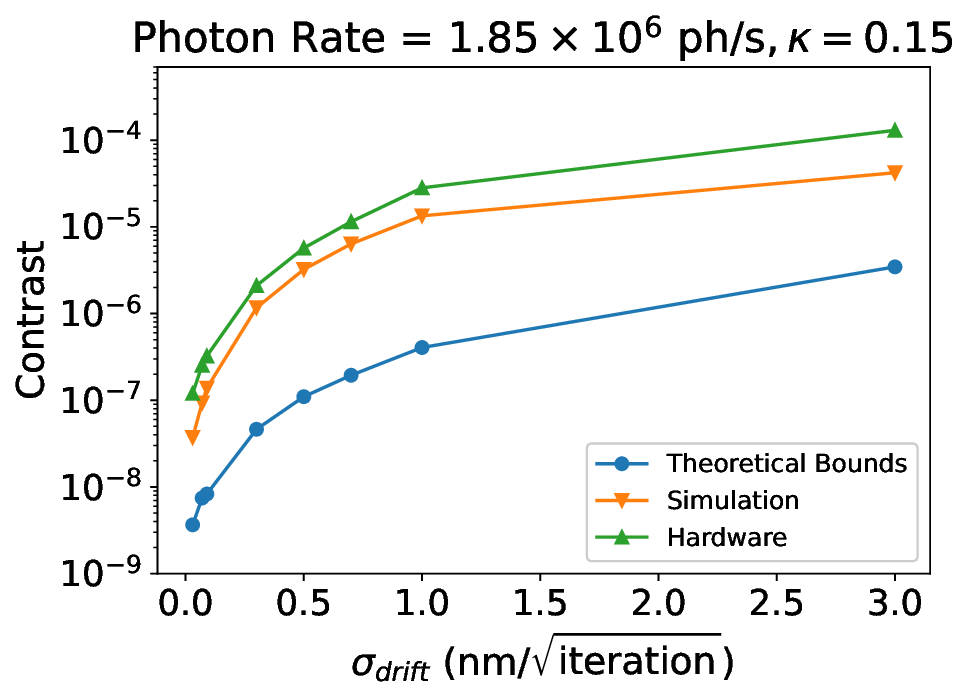}
    \end{subfigure}
    \hspace{0.05\textwidth}
    \begin{subfigure}[b]{0.4\textwidth}
        \centering
        \includegraphics[width=\textwidth]{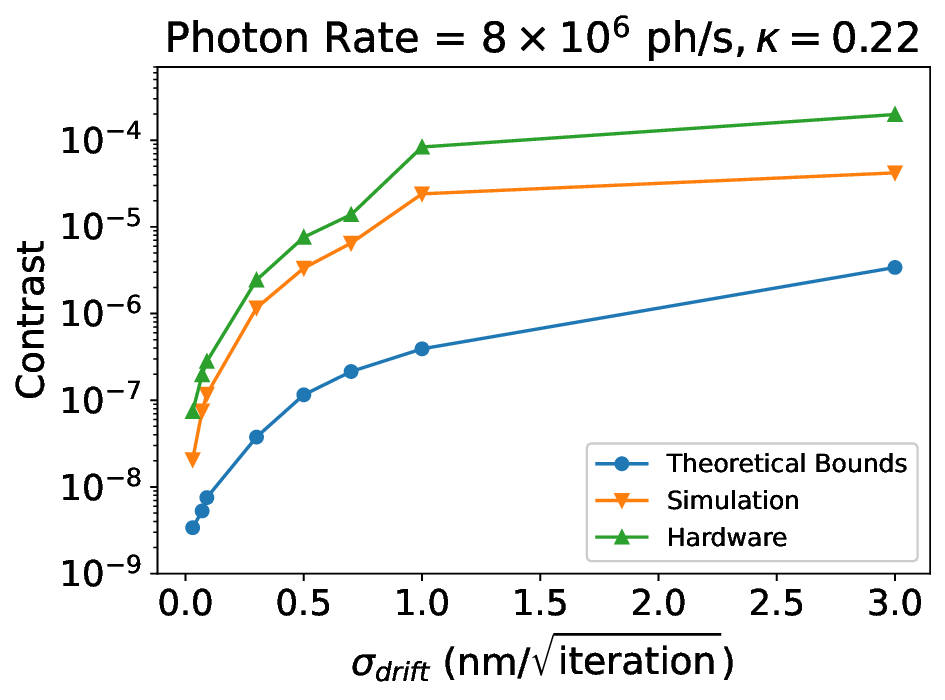}
    \end{subfigure}

    \vspace{0.03\textwidth}

    \begin{subfigure}[b]{0.4\textwidth}
        \centering
        \includegraphics[width=\textwidth]{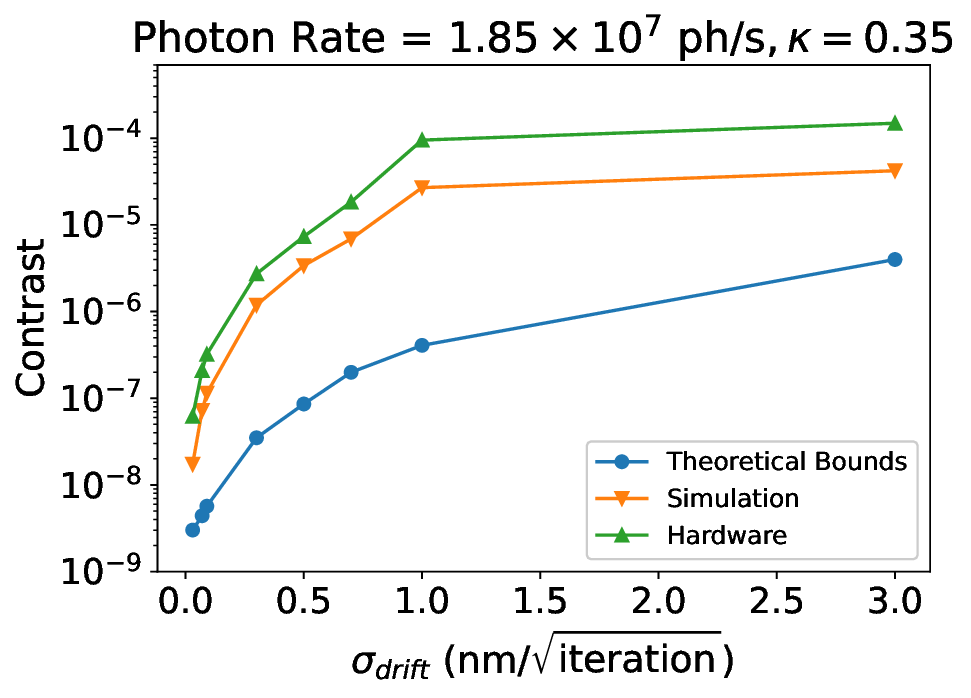}
    \end{subfigure}
    \hspace{0.05\textwidth}
    \begin{subfigure}[b]{0.4\textwidth}
        \centering
        \includegraphics[width=\textwidth]{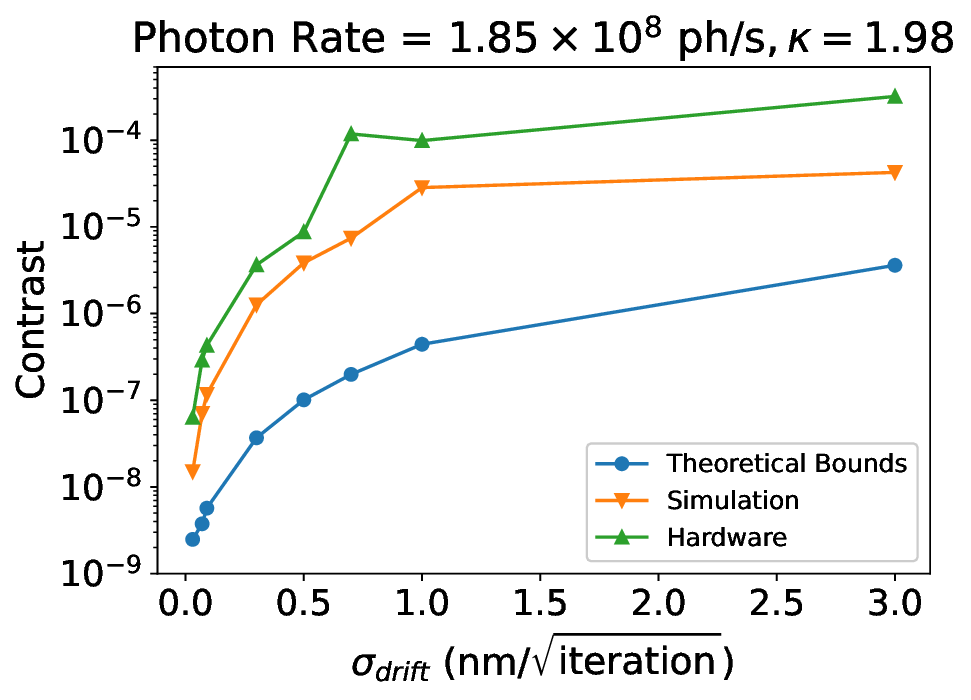}
    \end{subfigure}

    \caption{\label{fig:Combinede6} Mean contrast performance of theoretical bounds, hardware, and software runs at optimal dither for a direct photon rate of: Top left — $1.85 \times 10^{6}$~ph/s with $\kappa = 0.15$; Top right — $8.0 \times 10^{6}$~ph/s with $\kappa = 0.22$; Bottom left — $1.85 \times 10^{7}$~ph/s with $\kappa = 0.35$; Bottom right — $1.85 \times 10^{8}$~ph/s with $\kappa = 1.98$ \cite{manojkumar_parameter_2024}.}
\end{figure}

As shown in Figure~\ref{fig:Combinede6}, the HiCAT parameter scan revealed a substantial gap between theoretical bounds and achieved performance. Simulation results deviated from theory by a factor of $\sim$25, while hardware results diverged by a factor of $\sim$70, consistent across all photon rates tested. The discrepancy increased further at larger drift values, making the performance gap particularly pronounced in high-drift, high-dither regimes. These results highlight the need for more advanced algorithms to narrow the gap between theoretical limits and practical performance in contrast maintenance.

The remainder of this work is structured as follows: Section~\ref{sec:algo} introduces the Modal EKF algorithm; Section~\ref{sec:opt_dither} describes the stochastic gradient descent method for optimal dither selection; Section~\ref{sec: modal_paramscan} presents the parameter scan of the Modal EKF on the HiCAT simulator; Section~\ref{sec:param_results} discusses the results; and Section~\ref{sec:conclusion} provides conclusions and outlines future directions.

\section{MODAL EKF}\label{sec:Modal_EKF}
While the single-pixel EKF has demonstrated the ability to maintain contrast under photon noise and DM drift, its performance relative to the theoretical contrast bounds indicates room for improvement. The zonal approach of estimating the electric field at every science camera pixel does not fully capture the structure of the wavefront, which can instead be more accurately represented using a modal basis. As introduced by Pogorelyuk et al.~2022~\cite{pogorelyuk_dark_2022}, this modal architecture of estimation of wavefront error can be computed using a batch estimator, taking a collection of measurements, or sequentially using an EKF like DZM.

\subsection{Algorithm Formulation}
\label{sec:algo}
To reconstruct the open-loop electric field and the wavefront errors, an $r$-dimensional basis is chosen to represent the drifting electric field as a linear combination of the modal functions as 
\begin{equation}
\mathbf{E(u)} = \mathbf{J}\mathbf{u} + \mathbf{G}\mathbf{v} + \mathbf{E}_{\perp} \, ,
\end{equation}
where $\mathbf{J}\in\mathbb{R}^{2n \times m}$ is the Jacobian mapping $m$ DM commands, $\mathbf{u}\in\mathbb{R}^{m \times 1}$, to the science camera electric field with $n$ pixels, $\mathbf{G}\in\mathbb{R}^{2n \times r}$ maps the chosen modal basis to the field, $\mathbf{v}\in\mathbb{R}^{r \times 1}$ are the modal coefficients, and $\mathbf{E}_{\perp}\in\mathbb{R}^{2n \times 1}$ represents the remaining wavefront error that can not be defined by the DM effects or the chosen modal basis \cite{pogorelyuk_dark_2022}. Similar to pairwise probing, the difference between images taken with positive and negative probe commands, $\pm\delta\mathbf{u}_p$, is \begin{equation}
    \bm{I}(\delta\mathbf{u}_p) - \bm{I}(-\delta{\mathbf{u}_p}) = 4 \mathbf{B}\,\text{diag}(\mathbf{J}\delta\mathbf{u}_p)(\mathbf{J}\mathbf{u}+\mathbf{G}\mathbf{v}+ \mathbf{E}_{\perp}) \, ,
\end{equation}
with $\mathbf{B} \in \mathbf{R}^{n \times 2n}$ being the matrix that sums intensity values across channels, defined as $\mathbf{B} = \mathbf{I}\otimes\mathbf{1}$, where $\otimes$ is the Kronecker product, and $\mathbf{1}$ is a vector of ones. This difference in probe images provides information about the modal coefficients and serves as the EKF measurement.

If the modal basis is chosen to align with the DM basis, then $\mathbf{G}=\mathbf{J}$, and the state vector is defined as the modal coefficients $\hat{\mathbf{x}} = \hat{\mathbf{v}} \in \mathbb{R}^{m \times 1}$. The covariance $\mathbf{P}_0 \in \mathbb{R}^{m \times m}$ propagates with process noise
\begin{equation}
\mathbf{Q}_v = \sigma_{drift}^2 \mathbf{I},
\end{equation}
yielding $\mathbf{P}_{k+1} = \mathbf{P}_k + \mathbf{Q}_v$.

The observation matrix $\mathbf{H}_k \in \mathbb{R}^{n \times m}$ maps the probe-pair measurements, $\mathbf{y}(\pm \delta \mathbf{u}_p)$, to the state:
\begin{equation}\label{eqn: Modal Observation matrix}
\mathbf{H}_k = \frac{\partial(\mathbf{y}(\delta\mathbf{u}_p) - \mathbf{y}(-\delta\mathbf{u}_p))}{\partial \mathbf{v}_k} = 4\mathbf{B}\,\text{diag}(\mathbf{J}\delta\mathbf{u}_p)\mathbf{J}.
\end{equation}
The Kalman gain is then calculated
\begin{equation}\label{eqn: Modal Kalman gain}
\mathbf{K}_k = \mathbf{P}_k\mathbf{H}_k^T(\mathbf{H}_k\mathbf{P}_k\mathbf{H}_k^T + \mathbf{R}_p)^{-1},
\end{equation}
with the measurement noise covariance defined as 
\begin{equation}\label{measurement noise modal}
\mathbf{R}_p = \text{diag}\left(\mathbf{y}(\delta\mathbf{u}_p) + \mathbf{y}(-\delta\mathbf{u}_p) + 2(\mu_I t_{exp})^2 + 2\sigma_{rn}^2\right) ,
\end{equation}
where $\mu_I$ is dark current, $t_{exp}$ is exposure time, and $\sigma_{rn}$ is read noise.

Predicted measurements for the probe commands, $\pm\delta\mathbf{u}_p$, are calculated
\begin{equation}\label{eqn: predicted measurement}
\hat{\mathbf{y}} = \mathbf{\hat{y}}(\delta\mathbf{u}_p) - \mathbf{\hat{y}}(-\delta\mathbf{u}_p),
\end{equation} 
with \begin{equation}
\mathbf{\hat{y}}(\pm \delta\mathbf{u}_p) = \Re\left(\mathbf{J}(\mathbf{u} + \mathbf{\hat{v}} \pm \delta\mathbf{u}_p)\right) \circ \Re\left(\mathbf{J}(\mathbf{u} + \mathbf{\hat{v}} \pm \delta\mathbf{u}_p)\right) + \Im\left(\mathbf{J}(\mathbf{u} + \mathbf{\hat{v}} \pm \delta\mathbf{u}_p)\right) \circ \Im\left(\mathbf{J}(\mathbf{u} + \mathbf{\hat{v}} \pm \delta\mathbf{u}_p)\right) \, . \end{equation}
The state is then updated with the probe images, $\mathbf{y} = \mathbf{y}(\delta\mathbf{u}_p) - \mathbf{y}(-\delta\mathbf{u}_p)$, as
\begin{equation}\label{eqn: Modal EKF state update}
\hat{\mathbf{x}}_{k+1} = \hat{\mathbf{x}}_k + \mathbf{K}_k(\mathbf{y} - \hat{\mathbf{y}}),
\end{equation}
and the covariance is updated
\begin{equation}\label{eqn: Modal EKF covar update}
\mathbf{P}_{k+1} = (\mathbf{I} - \mathbf{K}_k\mathbf{H}_k)\mathbf{P}_k.
\end{equation}

The estimated modal coefficients $\hat{\mathbf{v}}$, estimating the wavefront error drift in the DM basis, serve as the control input, allowing the DM commands to be set directly as
\begin{equation}
\mathbf{u} = -\hat{\mathbf{v}},
\end{equation}
thus eliminating the need for EFC in the control loop.

\subsection{Optimal Dither Calculation}\label{sec:opt_dither}
As illustrated in Section~\ref{sec:contrast_limit_factors}, the probe command magnitude has a significant impact on estimator performance. While previous DZM studies identified optimal values by manually scanning across multiple dithers, it is possible to instead automate this process, allowing the algorithm to adaptively compute the optimal probe magnitude based on photon conditions and applied drift. The key source of information for this optimization is the covariance matrix $\mathbf{P}_k$, which quantifies the uncertainty in the modal coefficient estimates. By minimizing this uncertainty, one can identify a probe magnitude that provides sufficient phase diversity without excessively degrading contrast.

This formulation was introduced by Sun et al. 2020~  \cite{sun_efficient_2020} and defines the stochastic optimization problem as \begin{equation}\label{eqn: Stochastic optimization}
    \{\delta \mathbf{u}_k\} = \arg\min_{\{\delta \mathbf{u}_{k}\}}  \log \left| \mathbf{P}_k(\{\delta \mathbf{u}_k\}, \mathbf{x}_{k}) \right| 
+ \mathcal{P}(\{\delta \mathbf{u}_k\}) \, ,
\end{equation}
where $\mathbf{P}_k({\delta \mathbf{u}_k}, \mathbf{x}_{k})$ is the covariance matrix expressed as a function of the dither commands and the state estimate, and $\mathcal{P}({\delta \mathbf{u}_k})$ is a penalty function used to prevent excessively large probes. The penalty is defined in a Tikhonov-like form as
\begin{equation}\label{eqn: Stochastic penalty function}
\mathcal{P}({\delta \mathbf{u}_k}) = \alpha \delta \mathbf{u}_k^T \delta \mathbf{u}_k ,
\end{equation}
with $\alpha$ as the regularization parameter.

\begin{figure} [ht]
   \begin{center}
   \begin{overpic}[width=0.5\textwidth]{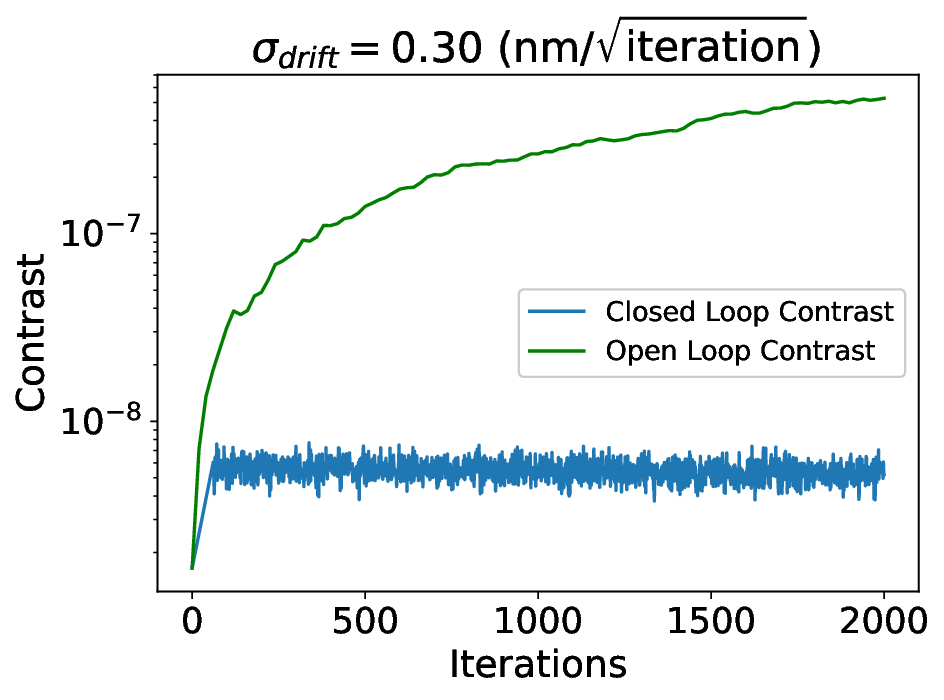}
    \put(23,20){\vector(0,1){5}}
    \put(22,17){\normalsize Optimal dither calculated and applied}
   \end{overpic}
   \end{center}
   \caption[example] 
   {\label{fig:opt_dither} 
Example run of the modal EKF with the optimal dither being applied at iteration 60, illustrating the drastic drop in contrast with the optimal dither being applied and the stability of contrast afterwards, as compared to the open-loop contrast.}
\end{figure}

The optimization is solved using stochastic gradient descent, where gradients of the covariance with respect to the dither commands guide iterative updates. This approach is implemented in Python using the PyTorch library~\cite{paszke_pytorch_2019}. Figure~\ref{fig:opt_dither} illustrates an example run of the modal EKF with the optimal dither calculation: after an initial convergence phase with a large dither (up to iteration 60), the optimal probe magnitude is calculated and applied, producing a sharp drop in contrast. The factor of $\sim$770 improvement between open-loop and closed-loop contrast performance demonstrates the efficacy of the algorithm.

\section{HICAT SIMULATIONS}
\label{sec:sections}
The modal EKF architecture was validated using the \texttt{catkit2} simulator for the HiCAT testbed, which was employed to develop the necessary algorithms and to generate comparable results prior to experimentation on the testbed.

\subsection{Modal EKF Parameter Scan}\label{sec: modal_paramscan}
\begin{figure} [ht]
   \begin{center}
   \begin{tabular}{c} 
   \includegraphics[width=0.95\textwidth]{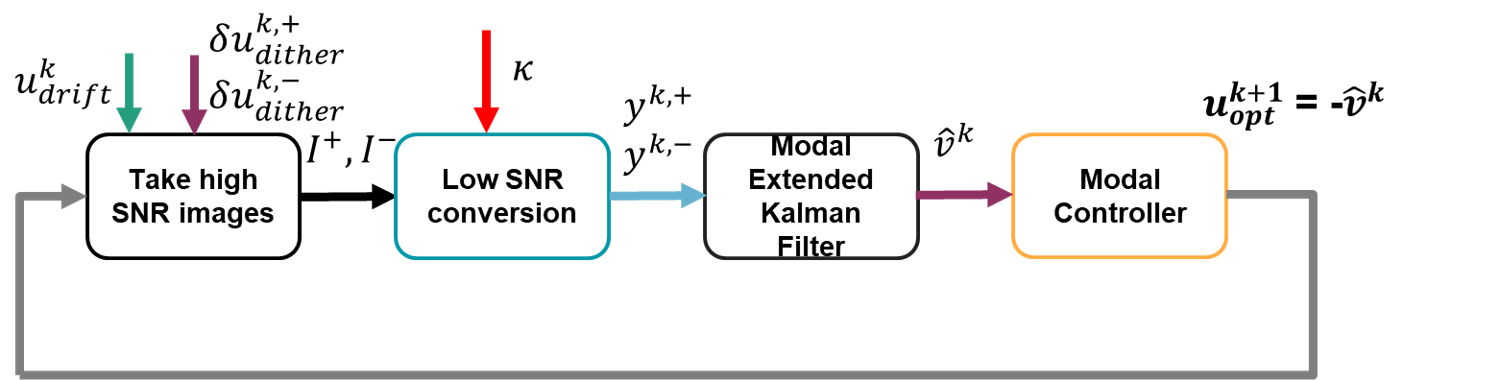}
   \end{tabular}
   \end{center}
   \caption[example] 
   {\label{fig:dzm_block_diagram} 
A block diagram of the parameter scan conducted using the modal EKF on the HiCAT simulator.}
\end{figure}

A modal EKF parameter scan was performed on the simulator, spanning the same parameters as in the DZM scan, namely the drift rate and the direct photon rate. The optimal dither implementation described in Section~\ref{sec:opt_dither} eliminated the need to span the dither parameter, as its value was instead computed during the experiment. The parameter scan process, illustrated in Figure~\ref{fig:dzm_block_diagram}, consisted of applying a drift and a positive–negative dither pair on the DMs. Each resulting image pair was passed through the low-SNR degradation process, determined by the photon flux parameter, and then provided as input to the modal EKF. The EKF’s estimates of the drifting modal components were negated to yield the optimal control command for the next iteration, thereby removing the need for EFC.

\begin{figure} [h]
   \begin{center}
   \begin{tabular}{c} 
   \includegraphics[width=0.5\textwidth]{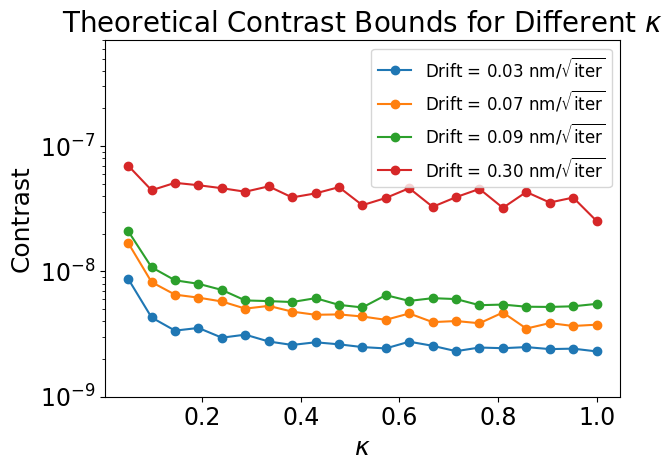}
   \end{tabular}
   \end{center}
   \caption[example] 
   {\label{fig:theory} 
Theoretical contrast curves for the drift rates of interest spanning various photon fluxes.}
\end{figure}

The drift rates used in the experiment matched those from the DZM HiCAT parameter scan, specifically the low-drift cases, as the higher rates were deemed unrealistic for the desired contrast levels. Photon values of interest were determined by generating theoretical boundary curves for the four drift rates of interest: $\sigma_{drift}=0.03, 0.07, 0.09,$ and $0.3 \; nm/\sqrt{iteration}$, across a range of photon fluxes, as shown in Figure~\ref{fig:theory}. These results revealed convergent contrast behavior near $\kappa=0.4$ photons per pixel in the DZ, marking the transition between the Poissonian and Gaussian regimes. This transition region was identified as particularly relevant, and photon fluxes in its vicinity were selected for the parameter scan. A summary of the chosen direct photon rates, mean photon values in the DZ, and drift rates is provided in Table~\ref{tab:photon_kappa_drift}.
\begin{table}[H]
\centering
\caption{Direct photon rates with corresponding initial mean number of DZ photons ($\kappa$) and drift magnitudes ($\sigma_{\text{drift}}$) tested on the Modal EKF parameter scan.}
\label{tab:photon_kappa_drift}
\begin{tabular}{|c|c|c|}
\hline
\textbf{Direct Photon Rate [ph/s]} & \boldmath{$\kappa$} & \boldmath{$\sigma_{\text{drift}}$} [nm/$\sqrt{\text{iteration}}$] \\
\hline
$5 \times 10^{7}$  & 0.32 & 0.03, 0.07, 0.09, 0.30 \\
$6.5 \times 10^{7}$ & 0.39 & 0.03, 0.07, 0.09, 0.30 \\
$8 \times 10^{7}$   & 0.44 & 0.03, 0.07, 0.09, 0.30 \\
\hline
\end{tabular}
\end{table}

\subsection{Parameter Scan Results}\label{sec:param_results}
\begin{figure}[htb]
    \centering
    \begin{subfigure}[b]{0.4\textwidth}
        \centering
        \includegraphics[width=\textwidth]{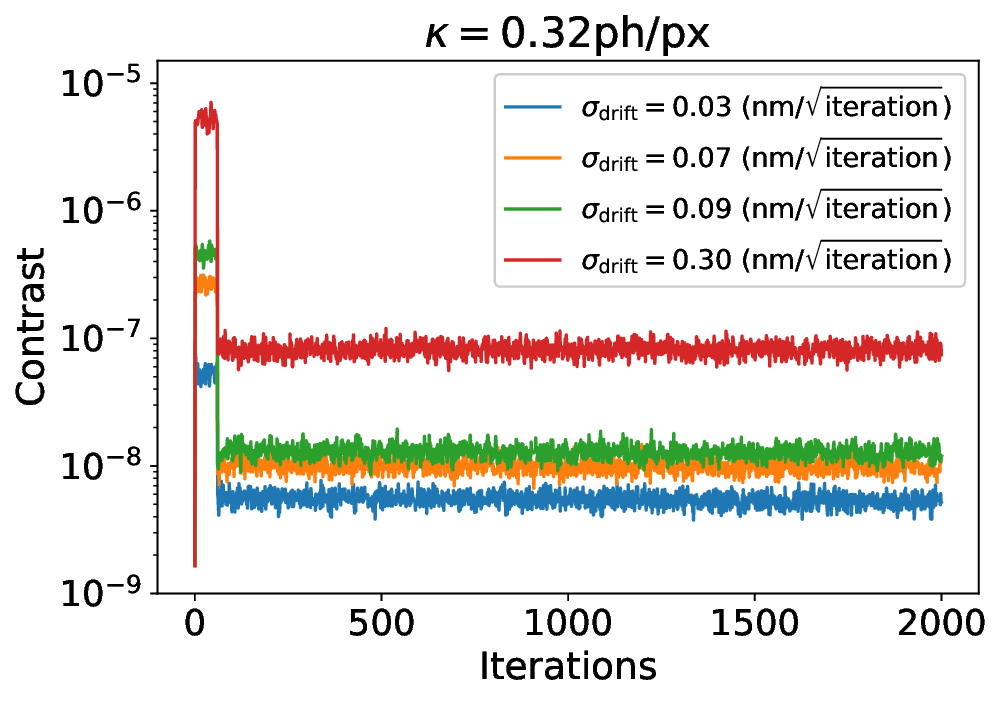}
    \end{subfigure}
    \hspace{0.05\textwidth}
    \begin{subfigure}[b]{0.4\textwidth}
        \centering
        \includegraphics[width=\textwidth]{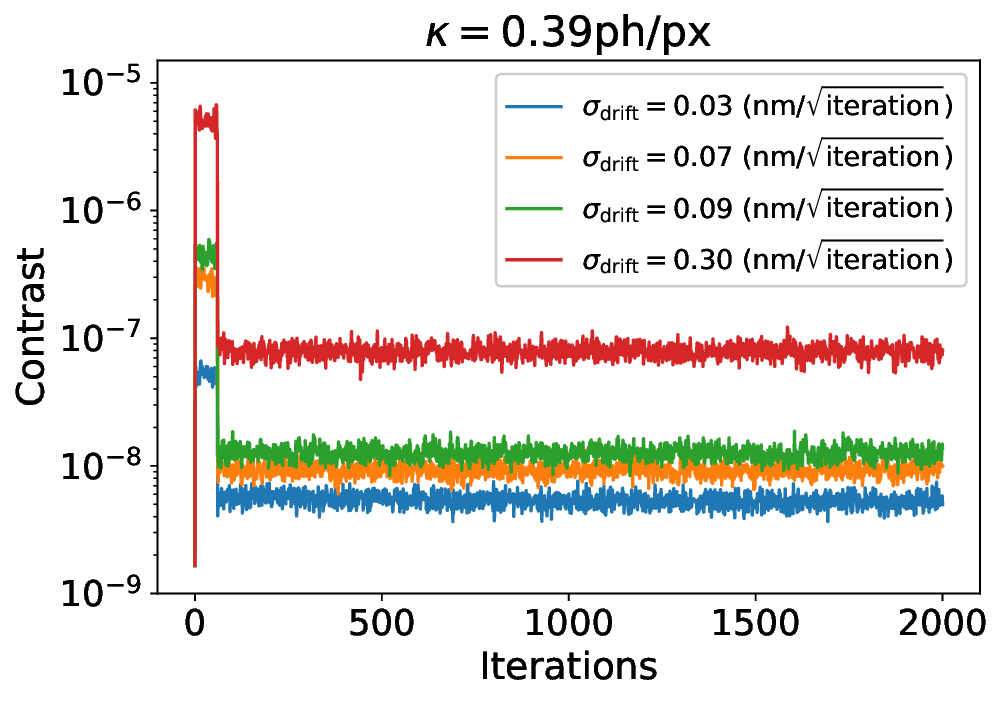}
    \end{subfigure}

    \begin{subfigure}[b]{0.4\textwidth}
        \centering
        \includegraphics[width=\textwidth]{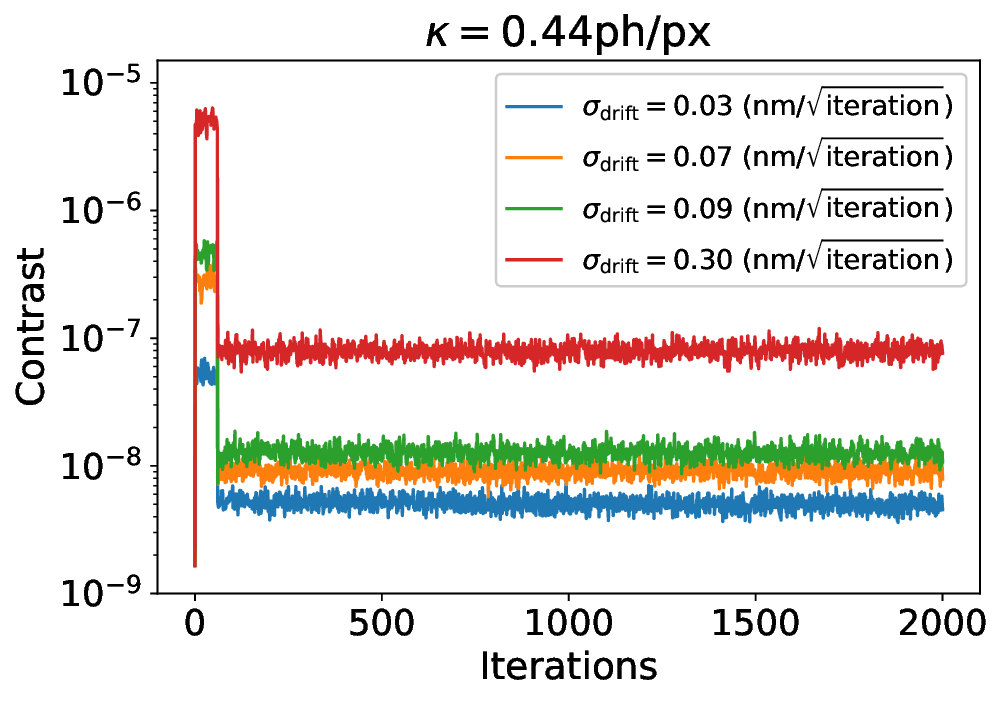}
    \end{subfigure}
    \hspace{0.05\textwidth}

    \caption{\label{fig:modal_history} Iteration history of all of the 12 various parameter combinations tested in the modal EKF parameter scan.}
\end{figure}

The parameter scan was conducted on the HiCAT simulator, \texttt{catkit2}, at a wavelength of~660 nm. Performance was evaluated across different drift rates and photon fluxes, with the optimal dither calculated at iteration 60. Each experiment was run for 2000 iterations, and results for the four drift rates and three photon fluxes are shown in Figure~\ref{fig:modal_history}. Across all 12 test conditions, the iteration histories demonstrated a sharp drop in contrast once the optimal dither was introduced, followed by stable contrast maintained for the remainder of the experiment. For all three photon conditions tested, only a slight improvement in contrast is observed at higher photon rates, whereas the contrast differences are more strongly driven by the drift magnitude, with lower drift values consistently yielding better performance.

\begin{figure}[htb]
    \centering
    \begin{subfigure}[b]{0.45\textwidth}
        \centering
        \includegraphics[width=\textwidth]{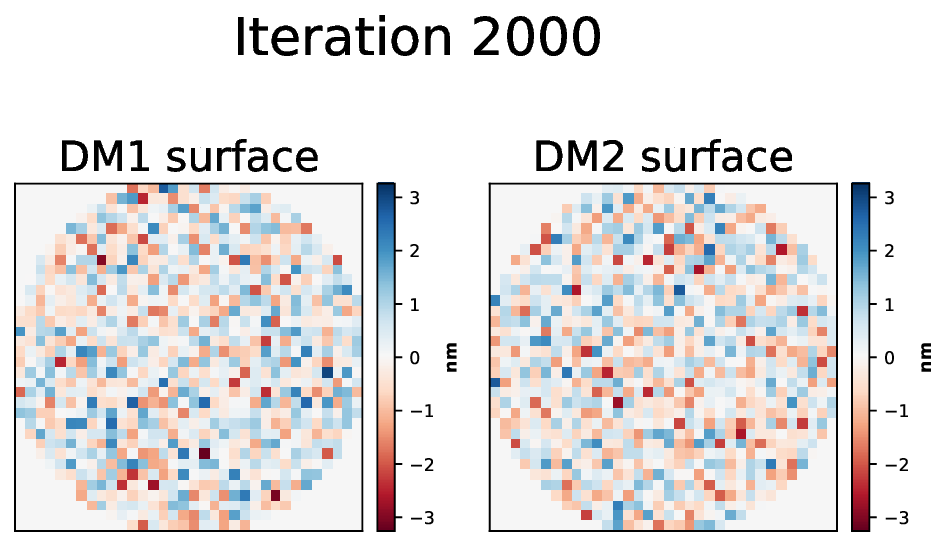}
    \end{subfigure}
    \hspace{0.05\textwidth}
    \begin{subfigure}[b]{0.45\textwidth}
        \centering
        \includegraphics[width=\textwidth]{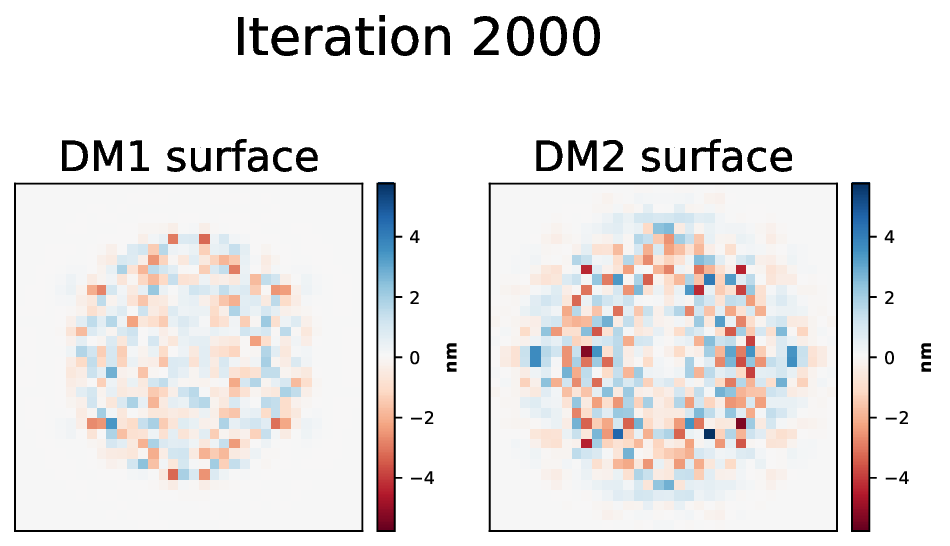}
    \end{subfigure}
    \caption{\label{fig:Drift_Est_imgs} Left: Diagram of drift commands on DMs; Right: Diagram of modal error estimates projected on DMs}
\end{figure}

When the modal EKF estimates are projected onto the DMs, they can be compared with the actual drifts applied during the experiment to assess algorithm performance. As shown in Figure~\ref{fig:Drift_Est_imgs}, the estimated drifting modes do not match the applied DM drifts, even though the EKF generates commands that successfully maintain contrast. This discrepancy reflects the nature of the two processes: while the DM drifts are randomly distributed across the entire DM surface, the EKF estimates are concentrated near the center and exhibit higher magnitudes. This suggests that the modal EKF prioritizes modes of greater significance, focusing its estimation on them rather than distributing effort across less impactful modes.

\begin{figure}[h]
    \centering
    
    \begin{subfigure}[b]{0.4\textwidth}
        \centering
        \begin{tikzpicture}
            \node[anchor=south west,inner sep=0] (image) at (0,0) 
                {\includegraphics[width=\textwidth]{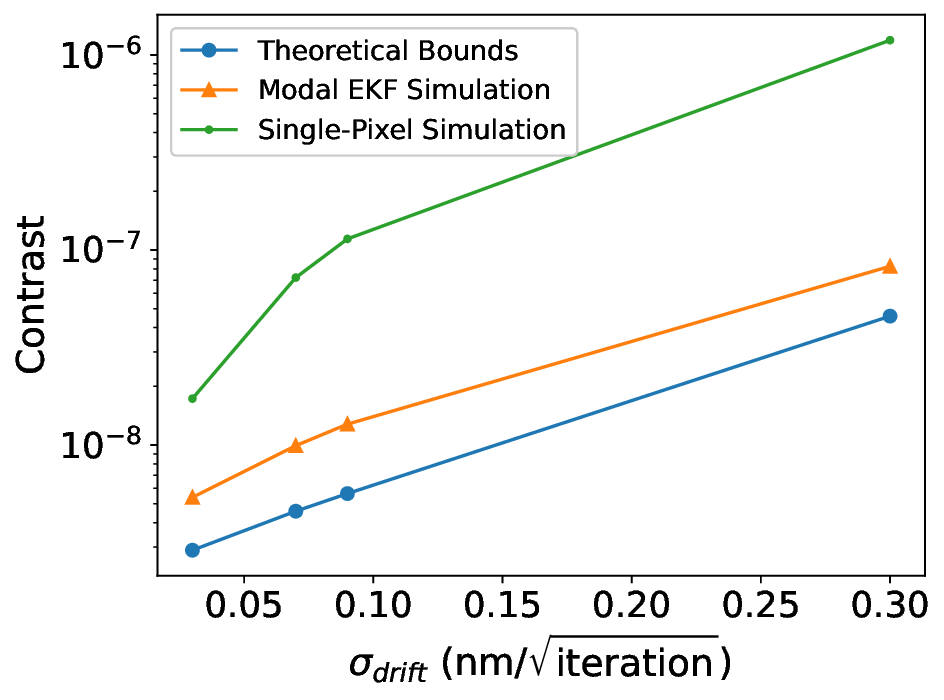}};
            \begin{scope}[x={(image.south east)},y={(image.north west)}]
                \draw[<->, thick] (0.96,0.55) -- (0.96,0.95);
                \node[anchor=west] at (0.84,0.8) {\small 26$\times$};
                \draw[<->, thick, cyan] (0.92,0.54) -- (0.92,0.63);
                \node[anchor=west, text = cyan] at (0.85,0.48) {\small 1.8$\times$};
            \end{scope}
        \end{tikzpicture}
        \caption{$\kappa = 0.32$}
    \end{subfigure}
    
    \begin{subfigure}[b]{0.4\textwidth}
        \centering
        \begin{tikzpicture}
            \node[anchor=south west,inner sep=0] (image) at (0,0) 
                {\includegraphics[width=\textwidth]{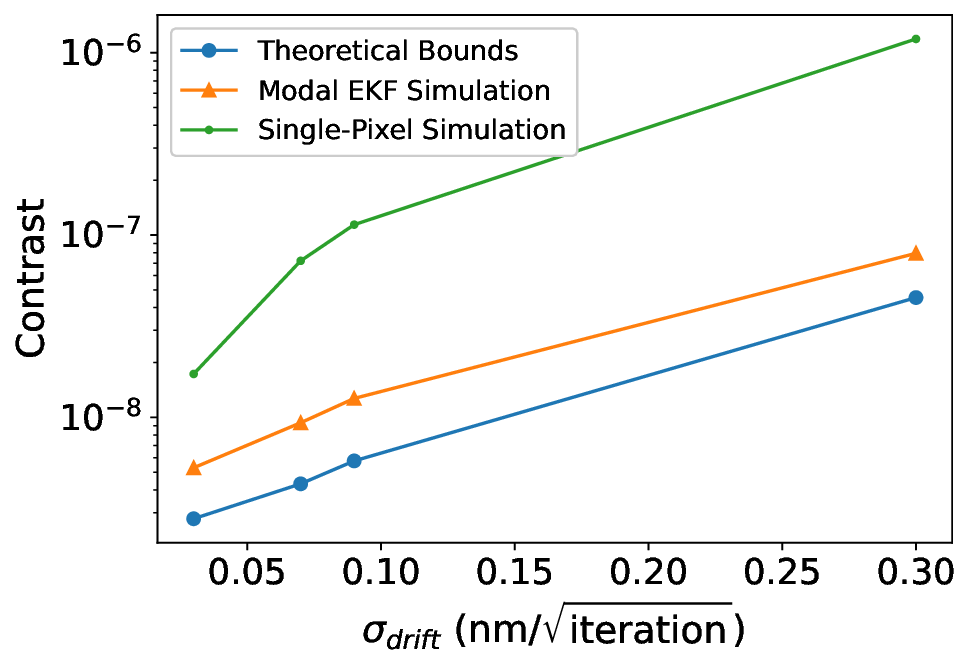}};
            \begin{scope}[x={(image.south east)},y={(image.north west)}]
                \draw[<->, thick] (0.96,0.55) -- (0.96,0.95);
                \node[anchor=west] at (0.81,0.8) {\small 26.2$\times$};
            \draw[<->, thick, cyan] (0.92,0.54) -- (0.92,0.63);
            \node[anchor=west, text = cyan] at (0.83,0.49) {\small 1.75$\times$};
            \end{scope}
        \end{tikzpicture}
        \caption{$\kappa = 0.39$}
    \end{subfigure}
    \vspace{0.5cm}
    \begin{subfigure}[b]{0.4\textwidth}
        \centering
        \begin{tikzpicture}
            \node[anchor=south west,inner sep=0] (image) at (0,0) 
                {\includegraphics[width=\textwidth]{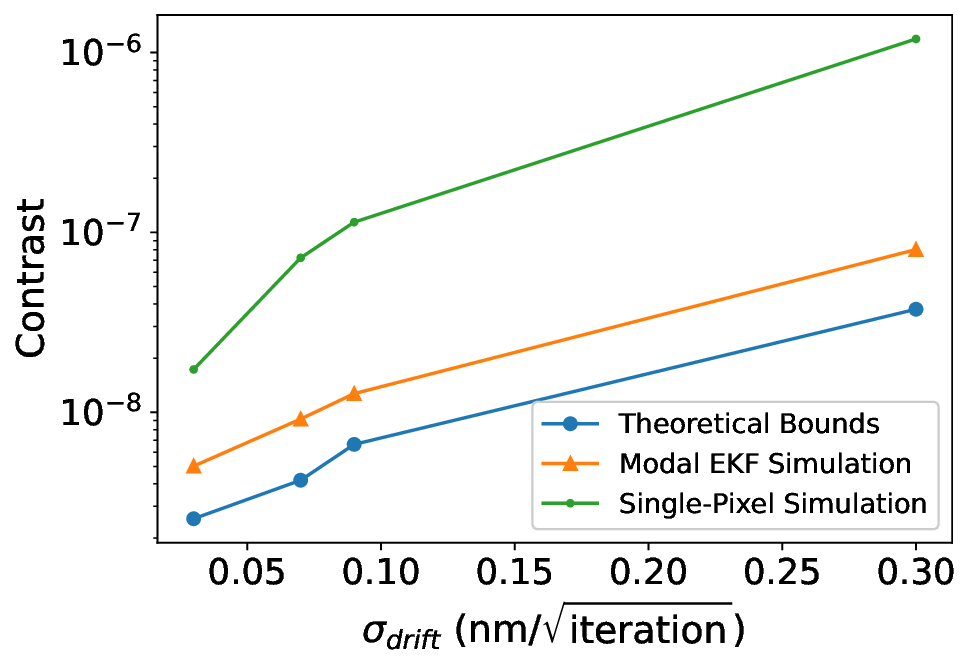}};
            \begin{scope}[x={(image.south east)},y={(image.north west)}]
                \draw[<->, thick] (0.96,0.53) -- (0.96,0.95);
                \node[anchor=west] at (0.81,0.8) {\small 31.8$\times$};
            \draw[<->, thick, cyan] (0.92,0.53) -- (0.92,0.63);
            \node[anchor=west, text = cyan] at (0.83,0.46) {\small 2.15$\times$};
            \end{scope}
        \end{tikzpicture}
        \caption{$\kappa = 0.44$}
    \end{subfigure}

    \caption{\label{fig:modal_compare} A comparison of the mean contrast achieved for the modal EKF parameter scan across various drift and photon rates, compared to the theoretical bounds and an example from the single-pixel EKF parameter scan from Manojkumar et al.\cite{manojkumar_parameter_2024} with a $\kappa=0.35$.}
\end{figure}

Finally, the mean contrast achieved with the modal EKF was compared against the theoretical bounds for each condition, as well as against results from the single-pixel EKF simulation on the \texttt{catkit2} simulator at comparable photon flux values. As shown in Figure~\ref{fig:modal_compare}, the modal EKF consistently achieved markedly better contrast across drift rates and photon fluxes. On average, its performance was within a factor of two to three of the theoretical bounds, whereas the single-pixel EKF deviated by as much as 30 times, representing nearly a 15-fold improvement relative to the standard DZM approach. The primary source of this improvement stems from the lower dither amplitude required by the modal EKF to sustain stable contrast, as opposed to the higher dither demands of the single-pixel EKF.

\section{CONCLUSION and FUTURE WORK}\label{sec:conclusion}
The simulation results of the modal EKF demonstrate its potential to significantly enhance the ability of coronagraph instruments to maintain high contrast in the presence of quasi-static DM drifts, outperforming the single-pixel EKF. For the HiCAT simulator, the modal EKF achieved up to a 15-fold improvement from the single-pixel EKF, relative to the theoretical bounds, underscoring the importance of future hardware testing to further validate the algorithm and raise the TRL of focal-plane WFSC maintenance methods.

While the results in Section~\ref{sec:param_results} were obtained on the in-air HiCAT testbed, which achieves a maximum simulated contrast of $10^{-9}$, it is also essential to evaluate the algorithm under conditions representative of HWO, namely a hardware $10^{-10}$ demonstration. This contrast level is achievable on the Decadal Survey Testbed (DST) at NASA JPL, located inside a vacuum chamber \cite{patterson_design_2019}.A preliminary parameter scan using the single-pixel EKF was performed and the results highlighted the strong dependence of contrast performance on dither amplitude and significant deviations from the theoretical bounds. These findings emphasize the need to identify optimal dither values and to reduce the contrast gap to within a factor of two to three of the theoretical limits.

The modal EKF, in its current formulation, estimates all drifting modes in the DM basis. However, depending on the near-singular structure of the Jacobian matrix, this full-state estimation can be computationally expensive and may introduce numerical instabilities. Moreover, as illustrated in Figure~\ref{fig:Drift_Est_imgs}, the EKF naturally converges on the most significant modes sensed in the focal plane, even when the drift is applied across the full DM surface. Hence, estimating a reduced set of dominant modes could reduce the computational cost, avoid numerical instabilities and decrease uncertainty by concentrating information into a smaller state vector.

To enable this, the Jacobian matrix $\mathbf{J}$ is decomposed using singular value decomposition (SVD):
\begin{equation}\label{eqn: SVD}
\mathbf{J} = \mathbf{U}\mathbf{\Sigma}\mathbf{V}^T ,
\end{equation}
where $\mathbf{U}$ contains the fundamental DM mode basis vectors, $\mathbf{\Sigma}$ is the diagonal matrix of singular values ordered by significance, and $\mathbf{V}$ is the orthogonal matrix of right singular vectors. The $r$ most significant modes are then selected by taking the first $r$ columns of $\mathbf{U}$, forming $\mathbf{G}_r \in \mathbb{R}^{2n \times r}$. The modal EKF then estimates the corresponding modal coefficients $\hat{\mathbf{v}}_r$, which can reconstruct the electric field as
\begin{equation}\label{eqn: Modal EKF Electric Field}
\delta\mathbf{E} = \mathbf{J}\mathbf{u} \approx \mathbf{G}_r\hat{\mathbf{v}}_r ,
\end{equation}
and derive the optimal DM command to null the electric field as
\begin{equation}
\mathbf{u} = -\mathbf{V}_r \mathbf{\Sigma}_r^{-1} \hat{\mathbf{v}}_r ,
\end{equation}
where $\mathbf{V}_r$ contains the first $r$ columns of $\mathbf{V}$ and $\mathbf{\Sigma}_r$ is the leading $r \times r$ block of $\mathbf{\Sigma}$.

Future work will focus on optimizing the modal EKF to estimate a reduced number of modes, with the goal of improving both its efficacy and numerical stability, validating in both simulation and hardware.

\appendix    

\acknowledgments 
This work was supported by the National Aeronautics and Space Administration under Grant NNH21ZDA001N issued through the Strategic Astrophysics Technology/Technology Demonstration for Exo-planet Missions Program (SAT-TDEM; PI: K. Cahoy). The research was carried out in part through the Jet Propulsion Laboratory, California Institute of Technology, under a contract with the National Aeronautics and Space Administration (80NM0018D0004). 
\bibliography{report} 
\bibliographystyle{spiebib} 

\end{document}